\documentclass{elsart}

\usepackage{graphicx}

\usepackage{amssymb}

\newcommand{\lm}{\mbox{lm}\,} 

\begin{document}

\begin{frontmatter}


\title{Generalized Additive Entropies in Fully Developed Turbulence}
\author[label1]{Patrick Ilg}
\ead{ilg@itp.physik.tu-berlin.de}
\address[label1]{
Inst. Theoret. Physics, TU Berlin, Sekr. PN 7-1, Hardenbergstr. 36, 
D-10623 Berlin, Germany}

\author[label2]{Iliya V. Karlin}
\ead{ikarlin@mat.ethz.ch}
\author[label2,label3]{Alexander N. Gorban}
\address[label2]{Department of Materials, Institute of Polymers, ML J19, 
ETH Z{\"u}rich, CH-8092 Z{\"u}rich, Switzerland}
\address[label3]{Institute of Computational Modeling RAS,
660036 Krasnoyarsk, Russia}
\ead{agorban@mat.ethz.ch}

\title{}


\author{}

\address{}

\begin{abstract}
We explore a possible application of the additive generalization 
of the Boltzmann-Gibbs-Shannon entropy proposed in 
[A.~N.~Gorban, I.~V.~Karlin, Phys.~Rev.~E, 67:016104 (2003)] 
to fully developed turbulence.
The predicted probability distribution functions are compared 
with those obtained from Tsallis' entropy and with experimental 
data. 
Consequences of the existence of hidden subsystems are illustrated. 
\end{abstract}

\end{frontmatter}

 
\section{Introduction}
In the past years, a body of experimental data   
of stationary statistical distributions has been collected 
that are not well described 
by the usual Boltzmann-Gibbs distribution 
(see e.g.~\cite{Abe2001} and references therein). 
In many cases, the 
concept of Tsallis entropy \cite{Tsallis88} together with the 
maximum entropy principle has been found useful to describe these data more 
accurately \cite{Abe2001}. 
The one-parametric family of Tsallis' entropies is defined as
\begin{equation} \label{S_q}
	S_q = \frac{1-\sum_ip_i^q}{q-1},
\end{equation}
where $q>0$ is the so-called nonextensivity parameter. 
If a system consists of two statistically independent subsystems 
then Tsallis' entropy of this system is not equal to 
the sum of the Tsallis' entropies of the subsystems for 
$q\neq 1$. 
The additivity is recovered only in the limit $q\to 1$, 
where Tsallis' entropy (\ref{S_q}) reduces to the 
classical Boltzmann-Gibbs-Shannon entropy $S_1$,
\begin{equation} \label{S_1}
	S_1 = \lim_{q\to 1}S_q = -\sum_i p_i\ln p_i. 
\end{equation} 
Since Tsallis' entropy is postulated rather than 
derived, this point is open to discussion 
\cite{Vives2002,KarlinGrmelaGorban2002}. 

Very recently, two of the present authors have derived a 
unique trace--form extension of the classical Boltzmann-Gibbs-Shannon entropy 
to finite systems that is still additive under joining 
statistically independent subsystems \cite{GK2003}, 
\begin{equation} \label{S_ast}
	S_\alpha^\ast = -(1-\alpha)\sum_ip_i\ln (p_i/p_i^0) + 
	\alpha\sum_i\ln (p_i/p_i^0), 
	\quad 0\leq \alpha \leq 1.
\end{equation}
In the limit $\alpha\to 0$, the 
Boltzmann-Gibbs-Shannon entropy is recovered, 
$S_1=S_0^\ast$.  
In Eq.~(\ref{S_ast}), $p_i^0$ denotes a general reference 
equilibrium state. In the sequel, we use equipartition as a reference, 
$p_i^0={\rm const.}$ 
Using a different parameterization, Eq.~(\ref{S_ast}) thus becomes
\begin{equation} \label{S_alpha}
	S_\alpha^\ast = -\sum_ip_i\ln p_i +\alpha\sum_i\ln p_i, 
	\quad \alpha \geq 0
\end{equation}
up to adding a constant and multiplying by a constant factor. 
It is shown in Ref.~\cite{GK2003} that the maximum entropy 
principle applied to this extensive generalized entropy 
results in non-exponential tails of probability 
distributions which are not accessible within 
classical Boltzmann-Gibbs-Shannon entropy  
(see also Refs.~\cite{GKOe2003} and \cite{PGorban2003}). 
The present paper provides an application of the additive
generalized entropies (\ref{S_alpha}) 
to experimental data on fully developed turbulence, 
where non-exponential tails of probability distributions 
have been observed \cite{Beck_PRE2001,Bodenschatz2001}.  
We also present consequences of incomplete description 
or hidden subsystems discussed in Ref.~\cite{GK2003}. 
Limitations of such an approach are discussed in 
Ref.~\cite{Kraichnan2003}.  

This paper is organized as follows. 
In Section \ref{entropies}, we briefly present the 
approach of Ref.~\cite{Beck_PRE2001} based on Tsallis' entropy 
and contrast it to
the alternative approach based on the additive entropies (\ref{S_alpha}).
In Section \ref{results}, we compare the results of 
the two approaches for describing the experimental data of 
Refs.~\cite{Beck_PRE2001,Bodenschatz2001}.


\section{Maximum Entropy and Generalized Probability Densities}
\label{entropies}
The maximum entropy principle under suitable constraints 
can be used to derive many 
relevant statistical distributions in physics 
but also in a wide range of other problems such as image reconstruction 
and time series analysis (see e.g.~A.~R.~Plastino in \cite{Abe2001} 
and references therein). 
Using only conserved quantities as constraints, the corresponding 
maximum entropy distributions describe equilibrium situations while 
the use of non-conserved quantities as constraints offers access 
to time-dependent processes. 

Consider a system with continuous state variable $u$, which is 
characterized by the energy $\epsilon(u)$. 
Let $p(u)$ denote the probability density of state $u$. 
Extremizing the entropy functional $S[p]$ subject to the constraint 
of fixed normalization $\int\!du\, p(u)=1$ and fixed total energy 
$\int\!du\, p(u)\epsilon(u)=E$, leads to 
\begin{equation} \label{maxEnt}
	\frac{\delta S[p]}{\delta p(u)} = \lambda_0 + \beta\epsilon(u),
\end{equation}
where $\delta/\delta p$ denotes the Volterra functional derivative and
$\lambda_0$ and $\beta$ are Lagrange multipliers that 
satisfy the constraints of fixed normalization and total 
energy, respectively. 
Thus, $\beta$ can be interpreted as a suitable inverse 
temperature. 
The solution to Eq.~(\ref{maxEnt}) gives the relevant 
or maximum entropy distribution consistent with the 
constraints. 

In case of Tsallis' entropy (\ref{S_q}), the solution to Eq.~(\ref{maxEnt}) 
are the probability distributions \cite{Tsallis98}
\begin{equation} \label{pBeck}
        p_q(u) = \frac{1}{Z_q}[1+\beta(q-1)\epsilon(u)]^{-1/(q-1)},
\end{equation} 
where $Z_q$ denotes a normalization constant. 
In the limit $q\to 1$, Eq.~(\ref{pBeck}) reduces to a Boltzmann 
factor $p_1(u)\propto e^{-\beta\epsilon(u)}$.


If instead of Tsallis' entropy (\ref{S_q}) 
the family of additive entropy functions 
(\ref{S_alpha}) is used,   
the maximum entropy condition 
(\ref{maxEnt}) becomes \cite{GK2003,GKOe2003} 
\begin{equation} \label{dS_alpha}
	\ln p_\alpha^\ast(u) -\alpha/p_\alpha^\ast(u) = -\Lambda(u),
\end{equation}
where $\Lambda(u)=1+\lambda_0+\beta\epsilon(u)$. 
Solving Eq.~(\ref{dS_alpha}) for $p_\alpha^\ast$ leads to 
\begin{equation} \label{pGK}
	p_\alpha^\ast(u) = \frac{\alpha}{\lm(\alpha e^{\Lambda(u)})}.
\end{equation}
In Eq.~(\ref{pGK}), use has been made of the modified 
logarithm $\lm y$, that denotes the solution to the transcendent 
equation $xe^x=y$. 
Note, that in the limit $\alpha\to 0$, the Boltzmann distribution 
is recovered from Eq.~(\ref{pGK}).

Due to their different analytical form, a direct comparison of 
the distribution functions (\ref{pBeck}) and  
(\ref{pGK}) is difficult. 
In the regime $\beta|\epsilon|\ll 1$, 
both, Eq.~(\ref{pBeck}) and Eq.~(\ref{pGK}) reduce to 
$p\propto (1-\beta\epsilon)$. 
For $\beta|\epsilon|\gg 1$, however, the distribution 
functions $p_q$ and $p^\ast_\alpha$ in general show a different 
behavior. 
While $p_q\propto (\beta\epsilon)^{-1/(1-q)}$ depends on the 
value of $q$ for $\beta|\epsilon|\gg 1$,
one finds a universal behavior  
$p^\ast_\alpha\propto (\beta\epsilon)^{-1}$, independent 
of the parameter $\alpha$ \cite{GK2003}. 
In the following section, some comparisons of 
Eqs.~(\ref{pBeck}) and (\ref{pGK}) are presented 
for a special choice of $\epsilon$.

Before proceeding to a specific example, we briefly 
address the problem of incomplete description as 
presented in Sec.~V.~of Ref.~\cite{GK2003}. 
Incomplete description in this context means, that 
in addition to $p$ other components or 
hidden subsystems $g(p)$ exist, 
whose entropy has to be taken into account. 
Define the two-parametric family of entropy 
functionals
\begin{equation} \label{S_alphat}
	S^\ast_{\alpha,t}[p] = (1-t)S^\ast_\alpha[p] + tS^\ast_\alpha[g(p)], 
	\qquad 0\leq t \leq 1. 
\end{equation}
In case of no hidden subsystems, $t=0$, Eq.~(\ref{S_alphat}) 
reduces to (\ref{S_alpha}), $S^\ast_{\alpha,t=0}=S^\ast_\alpha$. 
Applying the maximum entropy principle to the 
extended family of entropy functionals (\ref{S_alphat}), 
Eq.~(\ref{dS_alpha}) generalizes to 
\begin{equation} \label{dS_alphat}
	(1-t)\{ \ln p^\ast(u) - \alpha/p^\ast(u) \} + 
	   t \{ \ln g(p^\ast(u)) - \alpha/g(p^\ast(u)) \}J 
	= - \Lambda_t(u),
\end{equation}
where $\Lambda_t(u) = \Lambda(u) + t(J-1)$ 
and $J=\delta g(p^\ast)/\delta p(u)$. 
In particular, we consider the case 
\begin{equation} \label{ghost}
	g = 1 - \mu p, \qquad 0 \leq \mu \leq 1.
\end{equation}
The Fermi-Dirac entropy, for example, corresponds to 
$t=1/2$, $\alpha=0$, and $\mu=1$. 
Explicit solution of Eq.~(\ref{dS_alphat}) for $p^\ast$ is 
possible only for special cases. 

\section{Results for Fully Developed Turbulent Flows} 
\label{results}
The authors of Ref.~\cite{Beck_PRE2001} studied velocity
differences of high Reynolds number flow in a Taylor-Couette apparatus.
For sufficiently small distances, the experimentally observed
probability distribution function of the velocity differences
clearly shows non-exponential tails.    

In order to apply Eqs.~(\ref{pBeck}) and (\ref{pGK}) to this 
system, the expression for the energy $\epsilon$ has 
to be specified appropriately.  
In Ref.~\cite{Beck_PRE2001}, the energy $\epsilon(u)$ is assumed 
to be given by
\begin{equation} \label{eBeck}
        \epsilon(u) = \frac{1}{2}|u|^{2\zeta} -
        c\sqrt{\tau\gamma}\mbox{sgn}(u)(|u|^\zeta-\frac{1}{3}|u|^{3\zeta}),
\end{equation}
where the skewness $c$ and the intermittency parameter $\zeta$ are
related to Tsallis' $q$-parameter by
$c\sqrt{\tau\gamma}=0.124(q-1)$ and $\zeta=2-q$, respectively.
Thus, only one independent parameter $q$ is left. 
Beck offers in Ref.~\cite{Beck_PA2000} some arguments in favor of
Eq.~(\ref{eBeck}) for the case $\zeta=1$, while the extension to
$\zeta\neq 1$ in Ref.~\cite{Beck_PRE2001} is done in analogy 
to turbulence modeling.  
                                                            
For the distribution functions (\ref{pGK}), the parameter $\alpha$
needs to be specified. 
Remember, that the classical Boltzmann-Gibbs-Shannon entropy is 
recovered for $q\to 1$ in case of Tsallis' entropy (\ref{S_q}) and 
for $\alpha\to 0$ in case of the additive generalized entropy (\ref{S_alpha}). 
We suggest that the parameters $q$ and $\alpha$, describing the 
deviation from the classical Boltzmann-Gibbs-Shannon entropy 
are related by $\alpha=(q-1)^\nu$ with some exponent $\nu$. 
Since the parameters $q$ and $\alpha$ describe the 
non-ergodicity of the phase space dynamics, this relation might be 
interpreted in terms of excluded volume in phase space. 
Below, we use this simple power law relation as a plausible mapping
between the parameters of both the theories.

\begin{figure}[h]
        \setlength{\unitlength}{1cm}
        \begin{picture}(6,6)
        \put(0,0){\centerline{\includegraphics[width=10cm]{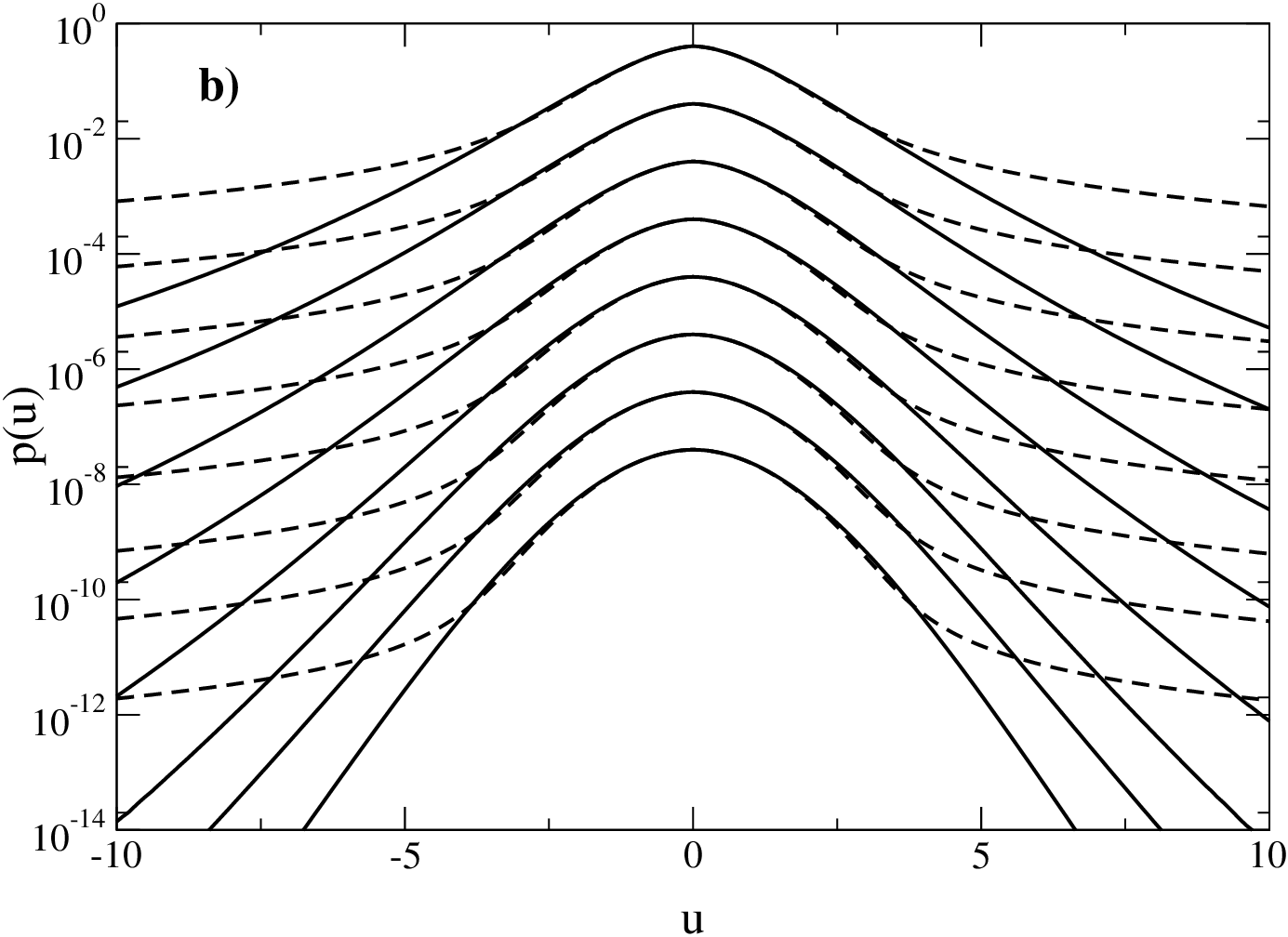}}}
\end{picture}
     \caption[] { \label{Fig_pu} 
	Probability distribution functions $p_q(u)$, 
	Eq.~(\ref{pBeck}), solid lines, and 
	$p^\ast_\alpha(u)$, Eq.~(\ref{pGK}), dashed lines: 
	(a) Linear plot, (b) logarithmic plot. 
	The values of the nonextensivity parameter $q$ are 
	from top to bottom: 
	$q = 1.168, 1.150, 1.124, 1.105, 1.084, 1.065, 1.055$ and 
	$1.038$, respectively. 
	These are the same values that are used in Ref.~\cite{Beck_PRE2001} 
	to describe experimental results of velocity differences. 
	For all curves the value $\alpha$ in Eq.~(\ref{pGK})  
	has been chosen as $\alpha=(q-1)^\nu$ with 
	$\nu=2.25$.}
\end{figure}

Fig.~\ref{Fig_pu} shows a comparison of the probability distribution 
functions (\ref{pBeck}) and (\ref{pGK}) with the energy given by 
(\ref{eBeck}). 
The values of the nonextensivity parameter $q$ are the same as 
used in Ref.~\cite{Beck_PRE2001}. 
For comparison with Eq.~(\ref{pGK}), exactly the same expressions 
for the energy, the parameters $\zeta$, $\beta$ and $c$ are used. 
We also choose exactly the same values for $q$ as done in 
Ref.~\cite{Beck_PRE2001}. 
Thus, the only parameter left is the exponent $\nu$, relating 
Tsallis' nonextensivity parameter $q$ to $\alpha$. 
The same value $\nu=2.25$ has been chosen in all cases to determine 
the parameter $\alpha$ in Eq.~(\ref{pGK}).   
On a linear scale, Fig.~\ref{Fig_pu} (a), the curves $p_q$ and 
$p^\ast_\alpha$ are almost indistinguishable by the naked eye. 
On a logarithmic scale however, differences between these 
curves are seen to become important for $u\gtrsim 4$. 
As mentioned above, the asymptotic behavior of $p^\ast_\alpha$, Eq.~(\ref{pGK}), 
is independent of $\alpha$, while the decay of $p_q$, Eq.~(\ref{pBeck}), can be 
varied by varying $q$. 
Thus, it appears the the distribution functions (\ref{pBeck}) 
describe the experimental data of Ref.~\cite{Beck_PRE2001} better 
than Eq.~(\ref{pGK}). 
We like to mention, however, that we made no attempt to improve 
the agreement of the distribution functions (\ref{pGK}) with 
Eq.~(\ref{pBeck}) by varying the relations between the parameters 
$\zeta$, $\beta$, $c$ and $q$. 

\begin{figure}[h]
	\vspace{1cm}
        \setlength{\unitlength}{1cm}
        \begin{picture}(6,6)
        \put(0,0){\centerline{\includegraphics[width=10cm]{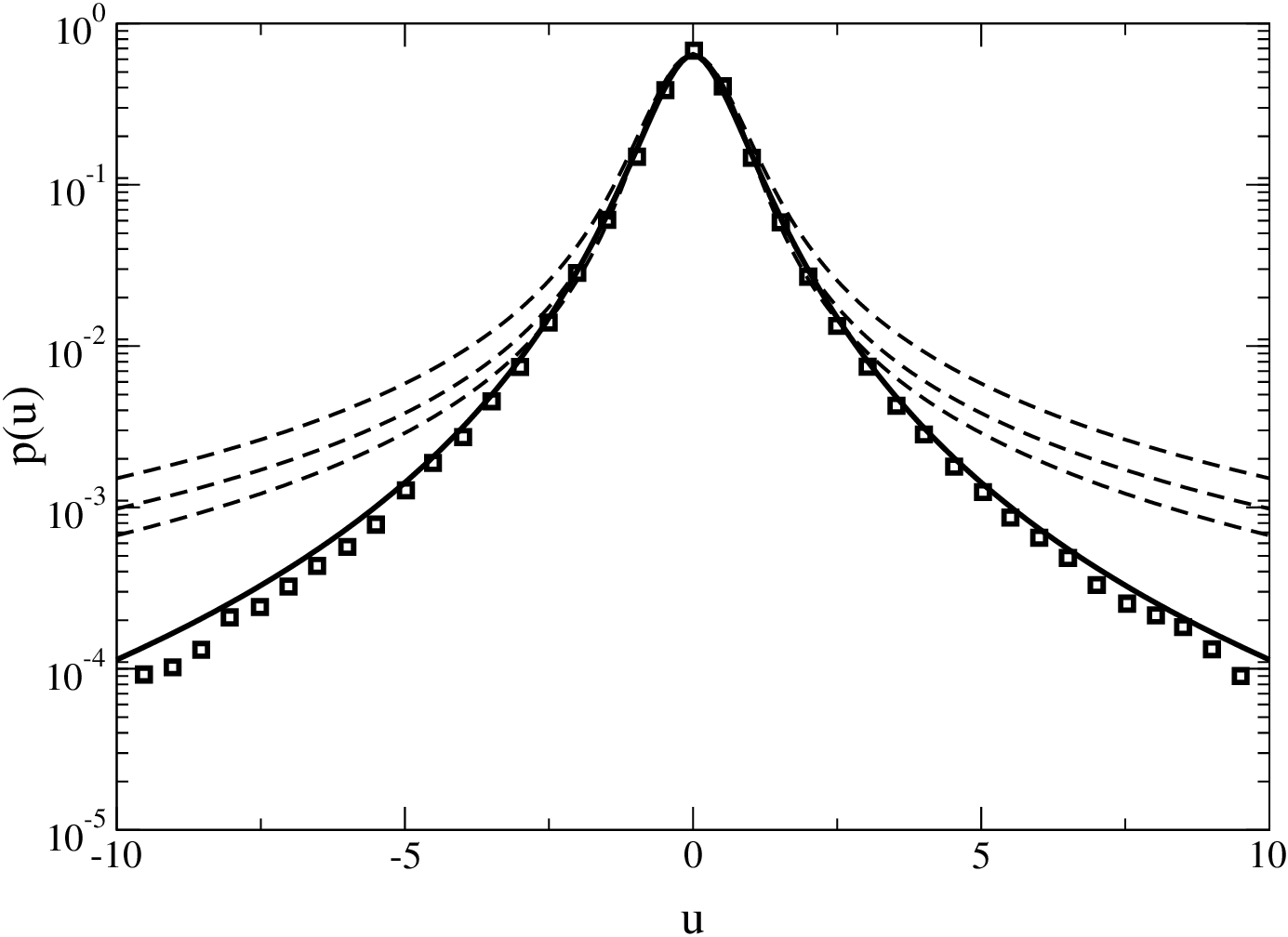}}}
\end{picture}
     \caption[] { \label{Fig_Bodenschatz} 
        Experimentally measured probability distribution of 
	normalized acceleration $u=a/\sqrt{\langle a^2\rangle}$ in Lagrangean 
	turbulence as measured by La Porta {\it et al.} \cite{Bodenschatz2001}, 
	symbols, and comparison with functions $p_q(u)$,
	Eq.~(\ref{pBeck}), solid line, and $p^\ast_\alpha(u)$, Eq.~(\ref{pGK}), 
	dashed lines. 
	The values of the parameters are $q=1.49$, $\zeta=0.92$, $c=0$ 
	for Eq.~(\ref{pBeck}). 
	The dashed lines correspond from top to bottom 
	to Eq.~(\ref{pGK}) with $\nu=2.25$,  
	Eq.~(\ref{dS_alphat}) with the same parameters and $t=\mu=0.5$ and 
	Eq.~(\ref{dS_alphat}) with $q=1.49$, $\zeta=1.0$, $c=0$, $\nu=2.25$ 
	and $t=\mu=0.5$.}
\end{figure}

In Ref.~\cite{Beck_PRL2001}, Beck provides a comparison of 
Eq.~(\ref{pBeck}) to the experimental results of La Porta {\it et al.}, 
\cite{Bodenschatz2001}. 
In the latter experiment, the acceleration of a test particle in 
a fully developed turbulent flow was measured. 
If the acceleration $a$ is interpreted as velocity difference on the 
smallest time scale of the turbulent flow (Kolmogorov time scale), 
the previous consideration apply also to this experiment with 
the identification $u=a/\sqrt{\langle a^2\rangle}$. 
Fig.~\ref{Fig_Bodenschatz} shows a comparison of the experimental 
results \cite{Bodenschatz2001} to the formulas (\ref{pBeck}) and 
(\ref{pGK}) with $q=1.49$, $\zeta=0.92$, $\beta=4$ and $c=0$, which are 
the values of the parameters proposed in \cite{Beck_PRL2001}. 
As noted in Ref.~\cite{Beck_PRL2001}, Eq.~(\ref{pBeck}) 
with this choice of parameters provides a very good description 
of the experimental results. 
The distribution function (\ref{pGK}) with the same values of 
parameters, however, overestimates the 
tails significantly already at $u\gtrsim 1$. 
The same value $\nu=2.25$ as before was chosen. 
Fig.~\ref{Fig_Bodenschatz} also shows the result of numerical 
solutions to Eq.~(\ref{dS_alphat}) for $t=\mu=0.5$ where all other 
parameters remain unchanged. 
Fig.~\ref{Fig_Bodenschatz} demonstrates that inclusion of a single 
hidden subsystem decreases the tails of the probability distribution. 
In the present case, this decrease leads to an improved comparison 
with the experimental data in a range $|u|\lesssim 3$. 
By including more hidden subsystems, a systematic improvement in the 
description of the experimental data is possible.

 
\section{Conclusions} \label{end}
We have presented an application of the additive generalization 
of the Boltzmann-Gibbs-Shannon entropy presented in Ref.~\cite{GK2003} 
to experimental data in fully developed turbulence 
\cite{Beck_PRE2001,Bodenschatz2001}. 
We found good agreement between the generalized distributions and 
the experimental data when compared on a linear scale. 
On a logarithmic scale, however, discrepancies in the tails of the 
distribution functions are evident. 
In particular, the generalized
distributions overpredict the tails in comparison with experiments. 
Improved experimental results presented very recently in 
Ref.~\cite{Bodenschatz2003b} seem to indicate that the tails of the 
probability distribution do not obey a power law behavior.
Thus, it appears that the maximum entropy distributions obtained 
either from Tsallis' or from the generalized additive entropy 
do not describe the tails of the distribution correctly, 
at least for the experimental results of 
\cite{Bodenschatz2001,Bodenschatz2003b}. 

It should be mentioned, that we used the same values of parameters 
that give very good agreement to the power-law distributions obtained 
from Tsallis' entropy and made no attempt to 
optimize this choice for the new distribution functions obtained from 
the generalized additive entropies.
Rather then trying to improve the fit of the experimental data 
by choosing different values of the parameters, we illustrate  
the consequences of incomplete description. We found that the inclusion 
of the entropy of a single hidden subsystem helps to improve the 
comparison to the experimental data significantly. 
Systematic improvements by including more hidden subsystems is 
straightforward.

\clearpage

%
\hspace{2cm}
\begin{figure}[h]
        \setlength{\unitlength}{1cm}
        \begin{picture}(6,6)
        \put(0,-1){\centerline{\includegraphics[width=10cm]{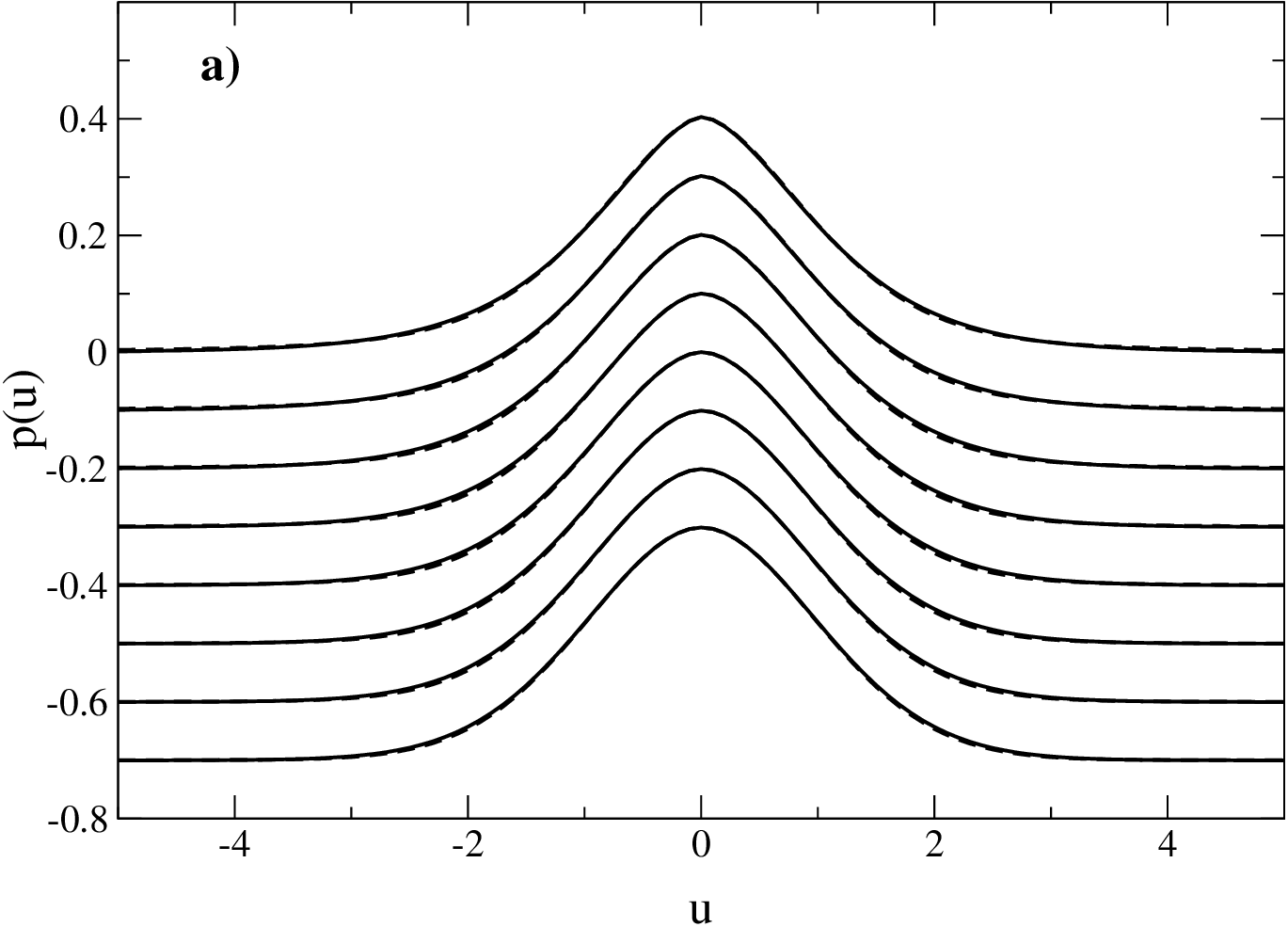}}}
\end{picture}
\end{figure}
\vspace{3cm}

\clearpage

%
%

\end{document}